\documentclass[reprint, amsmath,one column, amssymb, superscriptaddress, footinbib]{revtex4-2}

\usepackage{graphicx}
\graphicspath{{./plots f(r,t)/}}
\usepackage{amsmath}
\usepackage{natbib}
\usepackage{caption}
\usepackage{float}
\usepackage{dcolumn}
\usepackage{bm}
\usepackage[colorlinks=true,citecolor=blue,linkcolor=blue,urlcolor=blue]
{hyperref}
\usepackage{subfig}

\begin{document}
	
	\title{Investigating Evolving Wormholes in $f(R,T)$ Gravity}

	
	
	\author{Dhritimalya Roy}
	\email{rdhritimalya@gmail.com}
	\affiliation{Department of Physics, Jadavpur University, Kolkata-700032, India}
	
	\author{Ayanendu Dutta}
	\email{ayanendudutta@gmail.com}
	\affiliation{Department of Physics, Jadavpur University, Kolkata-700032, India}

	\author{Bikram Ghosh}
	\email{bikramghosh13@gmail.com}
	\affiliation{Department of Mathematics, Naba Barrackpore Prafulla Chandra Mahavidyalaya, Kolkata-700131, India}

	\author{Subenoy Chakraborty}
	\email{schakraborty.math@gmail.com}
	\affiliation{Department of Mathematics, Adamas University, Kolkata-700126, India}
	\affiliation{Research Fellow, INTI International  University, Putra Nilai-71800, Malaysia}

	
	\begin{abstract}

The present work examines whether evolving wormhole solution is possible or not in $f(R,T)$ modified gravity theory. In the background of inhomogeneous FLRW type wormhole configuration the field equations are investigated for different choices of  scale factors and shape functions. For the power law and exponential choice of the scale factor from cosmological context and decoupled power law of $f(R,T)$ in each variable, wormhole configuration has been examined for two viable choices of shape function. Energy conditions are examined graphically for a range of values of the parameters involved. Finally, the possibility of emergent scenario at early cosmic evolution has been examined.
	\end{abstract}

	\maketitle
	
	
	\section{Introduction}
	The concept of wormholes was first proposed by Einstein and Rosen, which is termed as Einstein-Rosen bridge\cite{Einstein:1935tc}. Wormholes act as a bridge between two distant points of the same universe or of different universes. The term wormhole was given by Misner and Wheeler\cite{Misner:1957mt}. Morris and Thorne were the first to put forward the concept of traversable wormholes\cite{Morris:1988cz}. They concluded that to sustain a traversable wormhole, the matter content at the throat needs to violate the Null Energy Condition(NEC). This violation is not possible with classical matter as they obey all the energy conditions, hence they proposed a matter called exotic matter which violates NEC. After their breakthrough paper many wormhole solutions were investigated by various researchers. The Ellis wormhole is one of the most common examples of wormhole solutions in General Relativity(GR)\cite{Ellis:1973yv, Ellis:1979bh, Visser:1995cc}. In the theoretical context, cosmology represents the most promising area for exploring the existence of exotic fluids. It is well established that the acceleration of the universe is linked to  matter, defined by the condition \(\omega < -\frac{1}{3}\) and governed by the equation of state \(p = \omega \rho\). Phantom energy, identified by a parameter \(\omega < -1\), exhibits unique characteristics, such as negative temperature and energy, where its energy density evolves alongside the expansion of the universe\cite{Caldwell:2003vq}.
	 To explain the expansion of the universe more precisely modified theories of gravity were developed and $f(R,T)$ theory of gravity\cite{Harko:2011kv} is one among many. Modified gravity theories allow for adjustment to the effective stress-energy tensor, offering an alternative approach to address the exotic matter problem while adhering to the energy conditions. Among the modified theories of gravity, $f(R,T)$ incorporates terms involving the  Ricci scalar and the trace of the energy momentum tensor in its  gravitational action. The inclusion of  the terms proportional to `$T$' is motivated by considering the quantum effects in gravity theory, or potential presence of imperfect fluids in the universe\cite{Harko:2011kv}. Hence consideration of these terms become particularly relevant when describing the matter content of wormholes, and are characterized by an perfect anisotropic fluid as well as imperfect fluids. These modified theories of gravity are now used extensively in wormhole physics to maintain the positivity of energy conditions and stands out as a crucial and intriguing challenge for availability of exotic matter. Various studies were made in different gravity theories(like $f(R)$ gravity theory,  Einstein-Gauss-Bonnet gravity,Lovelock theories ,etc) \cite{Lobo:2009ip, Mishra:2021xfl,Chakraborty:2007zi,Forghani:2019zbb,Maeda:2008nz,Dotti} to see whether they can limit the use of exotic matter. The  $f(R,T)$ gravity theory has already been shown to provide a good alternatives. Various authors have used $f(R,T)$ theory of gravity to obtain various wormhole solutions in the framework of $f(R,T)$ gravity\cite{ Moraes:2017mir, Yousaf:2017hjh,Elizalde:2018frj, Dutta:2023wfg}. Sahoo \textit{et.al}\cite{Sahoo:2017ual} investigated Wormholes in $R^2$-gravity within the $f(R,T)$ formalism. They concluded that the quadratic geometric and linear material corrections of this theory render the matter content of the wormhole remarkably able to obey the energy conditions. T.Azizi\cite{Azizi:2012yv}studied wormhole geometries in the context of $f(R,T)$ gravity by considering a particular EoS for the matter field account and they showed that effective stress-energy is responsible for violation of the NEC. Zubair  \textit{et.al}\cite{Zubair:2016cde} explored static spherically symmetric wormholes in $f(R,T)$ gravity and they revealed that the wormhole solutions can be constructed without exotic matter in few regions of space time. Moraes \textit{et.al}\cite{Moraes:2017rrv} investigated charged wormhole solutions in the $f(R,T)$ modified gravity. Phantom fluid wormholes are studied in $f(R,T)$ gravity\cite{Sahoo:2019ffc} by taking a EoS which is associated with phantom dark energy. Elizalde \textit{et.al}\cite{Elizalde:2019jeu} presented wormhole solutions using some wormhole models in $f(R,T)$ gravity and also examined all the energy conditions . In $f(R,T)$ gravity, spherical and hyperbolic wormholes have been studied by Zubair \textit{et.al}\cite{Zubair:2022jjm};  Lu\textit{et.al}\cite{Lu:2024qne} investigated the physical properties of traversable wormholes in this gravity.

	Non-stationary configurations can be achieved by introducing a time-varying metric, resulting in a dynamic wormhole setup. Visser\cite{Visser:1989kg} pioneered a category of traversable wormholes encompassing both static and dynamic types, employing a cut-and-paste technique (alongside appropriate Junction convergence criteria). Hochberg\cite{Hochberg:1990is} studied dynamic wormholes within higher-order $R^2$ gravity. Addressing the Horizon problem, other researchers\cite{Hochberg:1992du} utilized surgically modified dynamic wormhole solutions during the early stages of inflation in the universe. Several analogous solutions were explored by various authors (detailed references can be found in \cite{Roman:1992xj,Morris:1988tu}). The non-static Lorentzian wormhole model was proposed by Kar\cite{Kar:1994tz,Kar:1995ss}, demonstrating the feasibility of attaining a non-static wormhole geometry with throat matter that adheres to energy conditions. They further illustrated the possibility of a dynamic wormhole within FRW spacetime while satisfying all energy conditions. Subsequently, Hochberg and Visser\cite{Hochberg:1998ii} and Hayward\cite{Hayward:1998pp} demonstrated that the throat could be regarded as an anti-trapped surface, typically characterized by a violation of the NEC. 
	Numerous authors investigated evolving wormhole within higher-order gravity theories and within the framework of modified theories of  gravity\cite{Naz:2023pfl,Zubair:2022jjm,Saiedi:2021hkh,Zubair:2020uyb,Bhattacharya:2015oma,Mehdizadeh:2021kgv,Bhattacharya:2021frx,KordZangeneh:2020ixt,KordZangeneh:2020jio,Lobo:2020kxn,Mehdizadeh:2017dhb,Mehdizadeh:2012zz,SB,Yousaf:2019uky, Sefiedgar:2019xxd}.
	Extensive investigations into dynamic wormhole solutions supported by phantom matter were conducted by Cataldo \textit{et al.} \cite{Cataldo:2008pm,Cataldo:2008ku,Cataldo:2012pw,Cataldo:2013ala}, revealing that a dynamic wormhole could be sustained by a fluid mixture comprising multiple components, ensuring compliance with energy conditions. Maeda \textit{et al.} \cite{Maeda:2009tk} investigated a cosmological wormhole that satisfies the NEC, linking two FRW universes, wherein the entire spacetime is trapped, unlike scenarios where only the throat is trapped. Pan \textit{et al.} \cite{Pan:2014oaa} explored dynamic wormholes supported by two fluids in connection with particle creation mechanisms. 
	
	The traditional Big Bang cosmological model faces challenges concerning horizon and singularity similar to the event horizon of a black hole. As a result, there is a pursuit for cosmological frameworks that do not exhibit such problematic features. Ellis and Marteens \cite{Ellis:2002we} presented an alternate cosmological model, termed the Emergent Universe (EU), situated within the framework of Einstein's General Relativity, effectively tackling the singularity dilemma. Their proposition, stemming from Einstein's static universe, also resolves issues related to horizons. Since its origination, the Emergent Universe model has attracted significant interest from cosmologists, resulting in numerous investigations in this field \cite{Shekh:2023tbh,Sengupta:2023ysx,Shekh:2023rea,Ellis:2003qz}. Mukherjee et al. \cite{Mukherjee:2005zt,Mukherjee:2006ds} further expanded the Emergent Universe scenario to a flat spacetime, accommodating exotic matter that violates the energy conditions of General Relativity and elucidates the late-time acceleration of the universe. Various researchers have contributed to the development of the Emergent Universe model: Chakraborty \cite{Chakraborty:2014ora} explored particle creation within the framework of the Emergent Universe, Bose et al. \cite{Bose:2020xml} investigated the Emergent scenario within Hořava–Lifshitz gravity, and Bhattacharya and Chakraborty \cite{Bhattacharya:2016env} examined an Emergent Universe model within an inhomogeneous spacetime, drawing comparisons to the characteristics of Fred Hoyle's steady-state theory. Dutta et al. \cite{Dutta:2016kkl} expanded Mukherjee's equation of state to a more generalized form by examining modified Chaplygin gas in the Emergent Universe context under specific parameter conditions.

   Nevertheless, evolving wormholes have not been studied extensively in the $f(R,T)$ theory of gravity. Hence, motivated from above studies we intend to produce an exact solution of evolving wormholes in the framework of $f(R,T)$ gravity, and analyse its existence.
	The article is organized as follows: we begin with a brief review of the action and field equations followed by the geometry of the wormhole in the mentioned framework. Further, we investigate the energy conditions considering well known shape functions and use the power law and exponential form  of the scale factor. Lastly, we end the article with obtained results and discussions.

	\section{Basics of $f(R,T)$ gravity}\label{field_eq}
In the $f(R,T)$ gravity model, the action is expressed as\cite{Harko:2011kv}:
	\begin{equation}\label{eq1}
		S =\frac{1}{16\pi} \int d^4 x \sqrt{-g}  f(R,T)  + \int d^4 x \sqrt{-g}  \mathcal{L}_{m} ,
	\end{equation}
	where,$ f(R,T)$ represents an arbitrary function involving the Ricci scalar$(R)$, and $T$ denotes the trace term of the energy-momentum tensor.Here, $g$ stands for the metric determinant, $\mathcal{L}_{m} $ represents the matter Lagrangian density.

	Varying the action w.r.t the metric $ g_{\mu \nu} $, the field equation for $ f(R,T) $  theory of gravity is given by,
	\begin{eqnarray}\label{eq2}
		\nonumber	&&f_R(R,T) R_{\mu \nu} -\frac{1}{2} f(R,T) g_{\mu \nu}+(g_{\mu \nu} \Box -\nabla_\mu \nabla_\nu) f_R(R,T)\\ &&=  8\pi T_{\mu \nu} -f_T(R,T)(T_{\mu \nu}+\Theta_{\mu \nu}), 
	\end{eqnarray}
	where $ f_R(R,T) $ and $ f_T(R,T) $ are the differentiation of $ f(R,T) $ with respect to $ R $ and $ T $ respectively, and $ \Box f_R= g^{\mu \nu} \nabla_\mu \nabla_\nu f_R $. The variation of trace of energy-momentum tensor of the matter field, $ T= g^{\mu \nu} T_{\mu \nu} $ is given by:
	\begin{equation}\label{eq3}
		\frac{\delta(g^{\alpha \beta} T_{\alpha \beta})}{\delta g^{\mu \nu}} =T_{\mu \nu} +\Theta_{\mu \nu},
	\end{equation}
	where $ \Theta_{\mu \nu} $ and $ T_{\mu \nu} $ are given by
	\begin{eqnarray}\label{eq4}
		\Theta_{\mu \nu} \equiv g^{\alpha \beta} \frac{\delta T_{\alpha \beta}}{\delta g^{\mu \nu}},
		\label{eq5}\\
		T_{\mu \nu} \equiv g_{\mu \nu} \mathcal{L}_m - \frac{2 \partial\mathcal{L}_m}{\partial g^{\mu \nu}}.
	\end{eqnarray}
	
	Here, $ T_{\mu \nu} $ is chosen as energy-momentum tensor for the anisotropic fluid as:
	\begin{equation}\label{eq6}
		T_{\mu \nu}= (\rho+p_t)u_\mu u_\nu +p_t g_{\mu \nu} +(p_r-p_t)\chi_\mu \chi_\nu,
	\end{equation}
	where $ u_\mu $ is the timelike unit vector, $ \chi_\mu $ is a spacelike unit vector orthogonal to timelike unit vector, such that $ u_\mu u^\mu=-1 $, $ \chi_\mu \chi^\mu=1 $ and $ u_\mu \chi^\mu=0 $. Assuming the universal choice of the matter Lagrangian density $ \mathcal{L}_m=\rho$, we have $ \Theta_{\mu \nu}= -2T_{\mu \nu}+T g_{\mu \nu} $.

	Further, from Eq. \ref{eq6}, the field equation reads:
	\begin{equation}\label{eq7}
		G_{\mu \nu}=  8\pi G_{eff} T_{\mu \nu}+ T^{eff}_{\mu \nu} ,
	\end{equation}
	where
	\begin{eqnarray}
		G_{eff} &=& \frac{1}{f_R(R,T)} \left( 1+\frac{f_T(R,T)}{8\pi} \right), \label{eq8} \\
		\label{eq9}T^{eff}_{\mu \nu} &=& \frac{1}{f_R(R,T)} \Big[ \frac12 (f(R,T)-R f_R(R,T)  \\ \nonumber
		&+& 2\rho f_T(R,T)) g_{\mu \nu} -(g_{\mu \nu} \Box -\nabla_\mu \nabla_\nu) f_R(R,T) \Big].	
	\end{eqnarray}
	
	Note that, when  $ f_T(R,T)=0 $, we get  the usual $ f(R) $ theory of gravity . In these field equations, the energy-momentum tensor component ($ T_{\mu \nu} $) of $ f(R,T) $ gravity represents the interaction between matter and curvature, and one may interpret this as the curvature-matter coupling due to the exchange of energy and momentum between the both.The combined energy-momentum tensor has the form: 

		\begin{equation}\label{eq10}
	 T^{tot}_{\mu \nu}=T_{\mu \nu}+ \frac{1}{\left( 1+\frac{f_T(R,T)}{8\pi} \right)} T^{eff}_{\mu \nu}.
	\end{equation}

	However, $ f(R,T) $ gravity proposed by Harko \textit{et al.} \cite{Harko:2011kv} has a critical issue that violates the energy conservation law, and it leads to non-geodesic motion of test particles. But, an alternative approach of $ f(R,T) $ gravity proposed by Chakraborty \cite{Chakraborty:2012kj} showed that the form of the field equations remain similar if one takes into account the conservation of EM tensor. As a result, test particles move in geodesic orbits, and the choice of Lagrangian is not completely arbitrary. For a homogeneous and isotropic model of the universe, the approach refers to the field equations which is equivalent to the Einstein gravity with non-interacting 2-fluid system, one of which is the usual perfect fluid in modified gravity and the other one shows exotic nature.

	\section{The field equations and the wormhole geometry}

To solve the field equations and obtain a wormhole solution one may use the metric ansatz of the dynamical wormhole space-time which is given by\cite{Bhattacharya:2015oma} :
	
	\begin{equation} \label{eq11}
		ds^2 = -e^{{\Phi(r,t)} }dt^2 + a^2(t) \left [\frac{dr^2}{1- \frac{b(r)}{r}} + r^2 d{\Omega_2 }^2  \right],
	\end{equation}
	where, $ {\Phi(r,t)}$, is the redshift function, $a(t)$ is the scale factor of the wormhole universe,  $b(r)$ is the usual shape function of the wormhole , $d{\Omega_2 }^2 = d\theta ^2 + sin^2\theta d\phi^2$.
	
	The shape function present in the metric has to satisfy certain conditions and restrictions for equation (\ref{eq11}) to be a wormhole solution. The conditions imposed by Morris-Thorne are as follows\cite{Morris:1988cz}: 
	\begin{itemize}
		\item $b(r_0) = r_0$, where $r_0$ is the throat radius and is also a radial global minimum 
		\item  $b^\prime(r_0)<1 $ at $r=r_0$  and
		\item The Flaring out condition: $b(r) - r b^\prime(r) > 0 $
		\item To maintain traversability in a static wormhole configuration, one must also consider: the redshift function $\Phi(r)$ must be finite for $r >$$ {r_0}$ in order to avoid the presence of horizons and singularities. 
	\end{itemize}

Now, combining equations (\ref{eq7})-(\ref{eq9}) and the metric given by equation (\ref{eq11}), we get the field equations in the form as,

	\begin{widetext}
		\begin{eqnarray} \label{eqn13}
			3 e^{-2\Phi(r,t)} H^2 + \frac{b'}{a^2 r^2}  =  \frac{1}{f_R}\big[8\pi\rho-\frac{1}{2}f e^{2\Phi(r,t)} + \frac{1}{2} R  e^{2\Phi(r,t)} f_R + \rho f_T (1-e^{2\Phi(r,t)}) - 3 H \dot{f_R} + e^{2\Phi(r,t)} a^{-2}(1-\frac{b(r)}{r}) f_R'' \nonumber  & + & \\ e^{2\Phi(r,t) } f_R' (\frac{-3b - r {b\prime(r)} +4r}{2 a(t)^2 r^2})\big],
		\end{eqnarray}
		
		\begin{eqnarray}\label{eqn14}
	&	- e^{-2\Phi(r,t)} \left(\frac{2 \ddot{a}}{a} + H^2\right) - \frac{b}{a^2 r^3} + 2 e^{-2\Phi(r,t)} H \frac{\partial \Phi }{\partial t} +\frac{2}{a^2 r^2 }(r-b) \frac{\partial \Phi }{\partial r}\nonumber = \\ &\frac{1}{f_R}\big[8\pi p_r+p_r f_T  +  \left[{(f/2 - 3 \dot H f_R - 6 H^2 f_R - \frac{b'(r)}{a^2(t) r^2}f_R + \rho f_T)- e^{-2\Phi(r,t)}(\dot{f_R} \frac{\partial \Phi}{\partial t} - 2H\dot{f_R}- \ddot{f_R})}\right] \\ \nonumber
	&+ f_R' (\frac{\partial \Phi}{\partial r}+ \frac{2(r-b)}{a^2r^2} )\big],
		\end{eqnarray} 
		
		\begin{eqnarray}\label{eqn15}
			&	e^{-\Phi(r,t)} \left(\frac{2 \ddot{a}}{a} + H^2\right)  + \frac{b- rb'}{2 a^2 r^3} + 2 e^{-2\Phi(r,t)} H  \frac{\partial \Phi }{\partial t} + \frac{2 r- b- r b'}{2a^2 r^2 } \frac{\partial \Phi }{\partial r} + \frac{r-b'}{ a^2 r}\left[ \left( \frac{\partial \Phi }{\partial r}\right)^2 +\frac{\partial^2 \Phi }{\partial^2 r}\right]=\nonumber \\ 
			&\frac{1}{f_R}\big[8\pi p_t+p_t f_T  +   \left[{(f/2 - 3 \dot H f_R - 6 H^2 f_R - \frac{b'(r)}{a^2(t) r^2}f_R + \rho f_T + \frac{1}{2} f_R' \frac{rb'(r) - b(r)}{r(r-b(r))})- e^{-2\Phi(r,t)}(\dot{f_R} \frac{\partial \Phi}{\partial t} - 2H\dot{f_R}- \ddot{f_R})}\right] \nonumber \\ & - r^2 (1-\frac{b(r)}{r})[f_R'' + f_R' \frac{\partial \Phi}{\partial r}]- f_R' (b(r)-r) \big],
		\end{eqnarray}
		
		\begin{equation}\label{eqn16}
			2 \dot{a} e^{-\Phi(r,t)} \left(\sqrt{\frac{r-b(r)}{r}}\right) \frac{\partial \Phi(r,t) }{\partial r} = 0,
		\end{equation}
		
	\end{widetext}
	where, $ H = \frac{\dot{a}}{a}$ is the Hubble parameter, an `overdot' denotes the  differentiation w.r.t time $t$ and `prime'  denotes the differentiation w.r.t. to radial coordinate $r$,$ f_R $ and $ f_T$ are the differentiation of $ f(R,T) $ with respect to $ R $ and $ T $ respectively. Overdot on $ f_R $  denotes differentiation w.r.t $t$.

From the equation (\ref{eqn16}) it is observable that one may consider the redshift function, $\Phi$ to be independent of $r$. 

Here, we have chosen the redshift function, $\Phi=0$, without any loss of generality. Further, the reasons for this choice are as follows:
\begin{itemize}
	\item As the field equations in $f(R,T)$ gravity are very complicated, so to have a little bit simpler field equations  the redshift function, $\Phi =0 $ is assumed.
	\item $\Phi=0$ ensures that there is no horizon which is an important necessary condition for existence of wormholes.
	\item As we are considering wormhole geometry characterized by the shape function $b(r)$, the time dependence of $\Phi$ will not effect the wormhole geometry.
	\item The redshift function, $\Phi(r,t)=0$, gives a family of evolving wormhole conformally related to another family of zero tidal force evolving wormhole.
	\end{itemize} 
	
Thus for the evolving wormhole geometry , equations (\ref{eqn13})-(\ref{eqn16}) reduce to the simplified explicit form of the field equations as,
		\begin{eqnarray}\label{eqn17}
			3  H^2 + \frac{b'}{a^2 r^2} = \frac{1}{f_R}\big[8\pi\rho-\frac{1}{2} f + \frac{1}{2} R  f_R  - 3 H \dot{f_R} + a^{-2}(1-\frac{b(r)}{r}) f_R''+ f_R' (\frac{-3b - r {b\prime(r)} +4r}{2 a(t)^2 r^2})\big],
		\end{eqnarray}
		
		\begin{eqnarray}\label{eqn18 }
			-  \left(\frac{2 \ddot{a}}{a} + H^2\right) - \frac{b}{a^2 r^3}  = \frac{8\pi p_r+p_r f_T}{f_R}+ \frac{f}{2 f_R}+ \rho \frac{f_T}{f_R} - 3 \dot H - 6 H^2 -  \frac{b'(r)}{a^2r^2} + \frac{1}{f_R}\big[2H \dot{f_R}+\ddot{f_R}-\frac{2(r-b)}{a^2r^2} f_R'\big],
		\end{eqnarray} 
		
		\begin{eqnarray}\label{eqn19}
		&	- \left(\frac{2 \ddot{a}}{a} + H^2\right)  + \frac{b- rb'}{2 a^2 r^3}   =  \frac{8\pi p_t+p_t  f_T}{f_R}+ \frac{f}{2 f_R}+ \rho \frac{f_T}{f_R} - 3 \dot H - 6 H^2 -  \frac{b'(r)}{a^2r^2} + \\ \nonumber &\frac{1}{f_R}\big[\ddot{f_R}+2H\dot{f_R}-(\frac{1-\frac{b}{r}}{a^2}) f_R''- (\frac{-rb'+2r-b}{a^2r^2})f_R'\big].
		\end{eqnarray} 

	From the above field equations, one can obtain the energy-momentum tensor components, \textit{i.e} the energy density, the radial and transverse components of pressure. Here, we use decoupled power law form of $f(R,T)$, given by, $f(R,T)= \alpha R^m + \beta T$, where $\alpha$,$m$ and $\beta$ are constants. If one considers $\alpha,m=1$ and $\beta=2\lambda$, then it brings forth the very well known form of $f(R,T)$=$R+2\lambda T$. This known linear form was proposed by Harko\cite{Harko:2011kv} in their paper and is well established in $f(R,T)$ cosmology. As described in\cite{Roy:2022kid}, wormhole solutions in general relativity are constrained to be necessary with exotic matter. However, with a coupled geometric matter in terms of modified gravity can enable the possibility of non-exotic matter at the wormhole throat. The linear $f(R,T)$ theory is significantly simple and proposes a minimal modification on GR. Non-exotic traversable wormhole solutions in linear $f(R,T)$ gravity has been examined by Rosa and Kull\cite{Rosa:2022osy}, and they found the geometry satisfying all energy conditions in the entire spacetime. Therefore, in this article a coupling of geometric matter is considered to study its impact on the evolving wormhole geometry and on the energy conditions considering the above form.
	
	The Ricci scalar is given by:
	\begin{equation}\label{eqn24}
		R=6(\dot H + 2H^2) + \frac{2b'(r)}{a^2 r^2},
	\end{equation}
	
By separating the components of E-M tensor from the above obtained field equations, the generalised expressions for the energy density, and the pressure components are given by : 

	\begin{equation}\label{eqn20}
		\begin{aligned}
			 \rho = &\frac{1}{8 a(t)^2 r^2 (4 \pi + \beta) (8 \pi + \beta)} \Bigg[
			  b(r) (3 f_{R}' + 2 f_{R}'' r) (16 \pi + 3 \beta) 
			 + b'(r) (32 f_R \pi + 16 f_{R}' \pi r + 14 f_R \beta + 3 f_{R}' r \beta)\\
			  &+ r \Big(-4 f_{R}' (16 \pi + 3 \beta) - 2 f_{R}'' r (16 \pi + 3 \beta) 
			  + a(t)^2 r \Big(-6 \ddot{f_R} \beta + 2 R^m \alpha (8 \pi + \beta)   + 6 \dot{f_R} \dot{H} (16 \pi + 3 \beta) \\
			 &  + f_R \big(-16 \pi R + 6 \dot{H}^2 \beta - 5 R \beta + 48 \dot{H}^2 (2 \pi + \beta)\big)\Big)\Big)
			 \Bigg],
		\end{aligned}
	\end{equation}
	
\begin{equation}\label{eqn21}
	\begin{aligned}
p_r =& \frac{1}{8 a(t)^2 r^3 (4 \pi + \beta)(8 \pi + \beta)} \Bigg[ 
-b(r) \Big( 8 f_R (4 \pi + \beta) + r \Big( 64 f_{R}' \pi + 13 f_{R}' \beta
  - 2 f_{R}'' r \beta \Big) \Big) + r \Big( b'(r) \Big( f_{R}' r \beta + 2 f_R (16 \pi + \beta) \Big) \\
&  + r \Big( -2 f_{R}'' r \beta + 4 f_{R}' (16 \pi + 3 \beta)  - a(t)^2 r \Big( 2 R^m \alpha (8 \pi + \beta) + 2 \ddot{f_R} (16 \pi + \beta)  + 2 \dot{f_R} \dot{H} (32 \pi + 5 \beta)\\
&   - f_R \Big( 32 (3 \dot{H}^2 + \dot{H}^2) \pi + (2 \dot{H}^2 + R) \beta \Big) \Big) \Big) \Bigg],
\end{aligned}
\end{equation}


\begin{equation}\label{eqn22}
	\begin{aligned}
	p_t = &\frac{1}{8 a(t)^2 r^3 (4 \pi + \beta)(8 \pi + \beta)} \Bigg[ 
	16 b(r) \pi \Big( f_R - r \left( f_{R}' + 2 f_{R}'' r \right) \Big)
	 + b(r) \Big( 4 f_R - r \left( f_{R}' + 6 f_{R}'' r \right) \Big) \beta \\
	& + r \Big( b'(r) \Big( 16 f_R \pi - 16 f_{R}' \pi r - 2 f_R \beta - 3 f_{R}' r \beta \Big)  \quad + r \Big( 4 f_{R}' \left( 8 \pi + \beta \right) + f_{R}'' r \left( 32 \pi + 6 \beta \right) \\
	& \quad - a(t)^2 r \Big( 2 R^m \alpha (8 \pi + \beta) + 2 \ddot{f_R} (16 \pi + \beta)  + 2 \dot{f_R} \dot{H} (32 \pi + 5 \beta) - f_R \Big( 32 (3 \dot{H}^2 + \dot{H}^2) \pi + (2 \dot{H}^2 + R) \beta \Big) \Big) \Big) \Bigg].
	\end{aligned}
\end{equation}
		%
		%


	The above equations determines the relationship between the gravitational field and the matter field of the evolving wormhole. These equations of the components of the energy momentum tensor allow us to form the exact solutions of the matter content of the wormhole by considering specific form of the shape function and the scale factor.
	
\begin{figure*}[h!]
		\centering
		\subfloat[]{{\includegraphics[width=8cm]{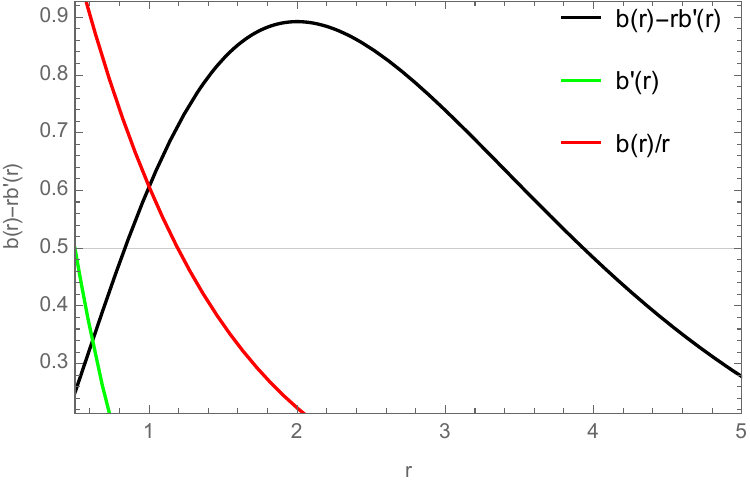}}}\qquad
		\subfloat[]{{\includegraphics[width=8cm]{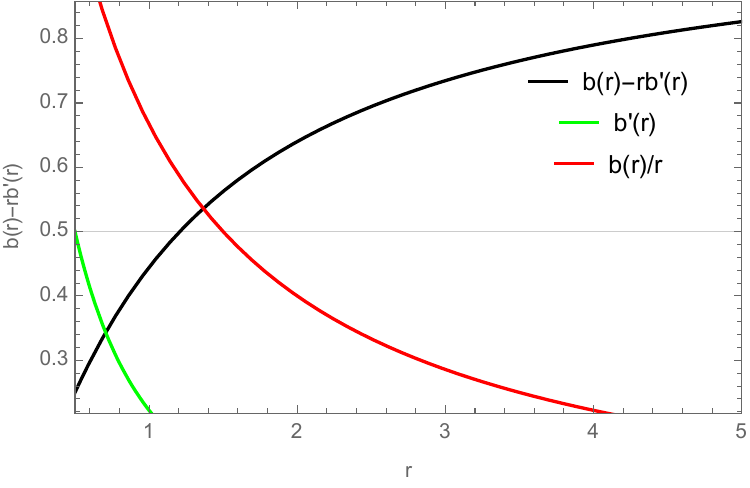}}}\qquad
		\subfloat[]{{\includegraphics[width=8cm]{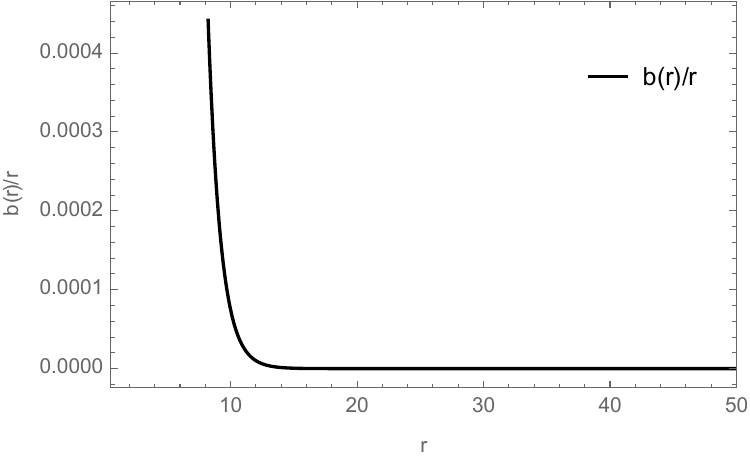}}}\qquad
		\subfloat[]{{\includegraphics[width=8cm]{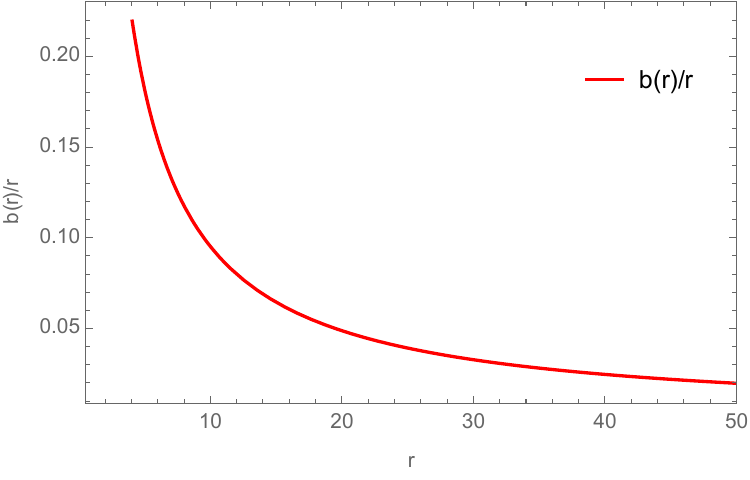}}}	
		
		\caption{Plots demonstrating the Morris-Thorne conditions for the shape function with radial distance $ r $, a) for Model I and b) for Model II. Here ($c$) and ($d$) shows the asymptotic behaviour of the shape functions for the Models I and II respectively The nature of the plots are obtained considering $ r_0=0.5 $(throat~ radius).}
		\label{shapefunction}
	\end{figure*}

Here,  we use two different well known shape functions given as\cite{Ghosh:2022vit, Ghosh:2021tml}:
	\begin{equation}\label{eqn26}
		b(r)= r/(e^{r - r_0})~~{(Model ~I)},
	\end{equation}
	\begin{equation}\label{eqn27}
		b(r)=r/(1 + r - r_0)~~{(Model~ II)},
	\end{equation}
	where $r_0$ is the throat radius. 
The  shape functions used here provide a valid wormhole solution satisfying the Morris-Thorne condition. The plots provided in Figure \ref{shapefunction} show that the Models used here satisfy conditions that are required for it to be a shape-like function. 
	
	For the scale factor, we consider the two forms:
	\begin{itemize}
		\item Power law form:
		\begin{equation}\label{eqn23}
			a(t)=a_0 t^n,
		\end{equation}
		\item Exponential form:
		\begin{equation}
			a(t) = [a_0+ e^{\mu t^n}]\label{sf eu},
		\end{equation}
	\end{itemize}
	with $a_0$, $\mu$ and $n$ as constants. The above two choices of the scale factor are nothing but corresponds to power law form of expansion and exponential rate of expansion respectively. The motivation for these two choices is from cosmology where the expansion of the universe is assumed to be any one of these two forms, depending on the cosmic scenario.

	We will now use the above two shape functions for two different choices of scale factors for analysing the energy conditions.

%
	
	\section{Energy Conditions}\label{en_con}
	
In the Einstein's gravity, for the existence of traversable wormholes NEC must be violated at the throat. Notably, certain theories within this framework support types of matter that adhere to energy conditions in diverse manners. In a recent study\cite{Ilyas:2022zoq}, it was demonstrated that within the framework of $ f(R,T) $ gravity, a wormhole can emerge using non-exotic matter, with the conventional EM tensor satisfying the null energy condition (NEC). The additional curvature induced by the modified gravity alters the NEC violation, resulting in an overall violation of the NEC by the total EM tensor. A broader discussion on the potential existence of wormholes with ordinary matter in modified gravity is outlined in the paper  \cite{Roy:2022kid}. Another instance of a wormhole with non-exotic matter in the context of $ f(R,T) $ gravity is presented in the work by Banerjee et al. \cite{Banerjee:2020uyi}. The above mentioned studies were however all done on static wormhole geometry.

	 In this article we investigate whether an evolving wormhole can exist in $f(R,T)$ gravity by satisfying the energy conditions.
	The energy conditions, namely the Null Energy Condition (NEC), Weak Energy Condition (WEC), Dominant Energy Condition (DEC), and Strong Energy Condition (SEC), are obeyed when the following inequalities are satisfied:
	
	\begin{itemize}
		\item (i)NEC: $\rho+{(p_r)} \geq 0$,  $\rho+{(p_t)} \geq 0;$
		\item (ii)WEC:$\rho \geq 0$, $\rho+{(p_r)} \geq 0$, $\rho+{(p_t)} \geq 0$;
		\item (iii)SEC:$\rho+{(p_r)} \geq 0$,  $\rho+{(p_t)} \geq 0$ and $\rho+{(p_r)} +2{(p_t)} \geq 0$;
		\item (iv)DEC:$\rho \geq 0$, $\rho-|{(p_r)}| \geq 0$, $\rho-|{(p_t)}| \geq 0$.
	\end{itemize}

From the above quantities we now plot the energy conditions and see whether they are satisfied or not. The choices of the scale factor is given by equations (\ref{eqn23}) and (\ref{sf eu}), and the shape functions  are given by equations (\ref{eqn26}) and (\ref{eqn27}). 

In the expressions there are seven parameters namely, $a_0,~r_0,~\alpha,~\beta,\mu,m~and~n$. The values of $a_0~and~r_0$ are somewhat arbitrary as they can be used for rescaling the solutions without loss of generality. The different models proposed here are as follows:
 \\
 
$\bullet$ \textbf{Power Law Scale factor:}
 
\begin{figure*}[h!]
	\centering
	\subfloat[]{{\includegraphics[width=8cm]{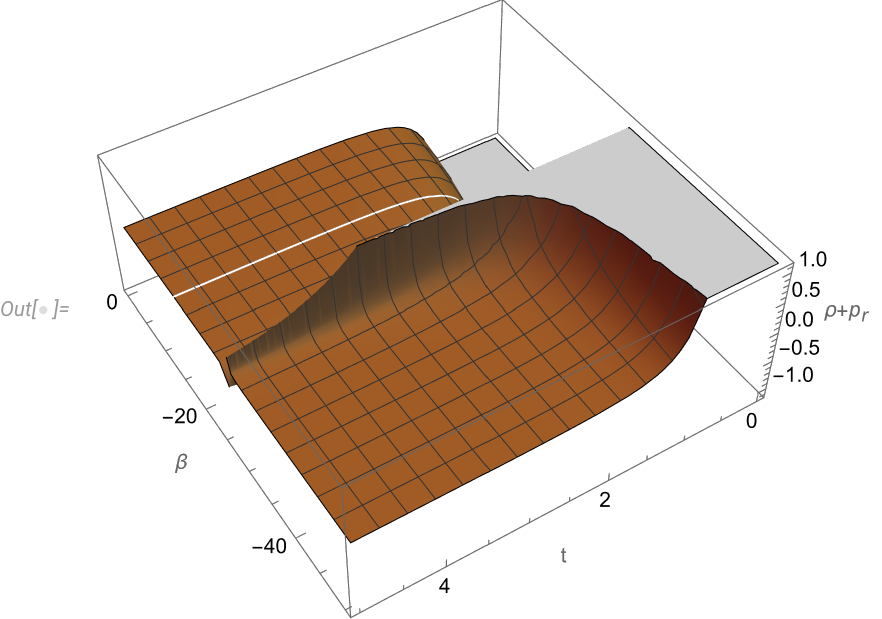}}}\qquad
	\subfloat[]{{\includegraphics[width=8cm]{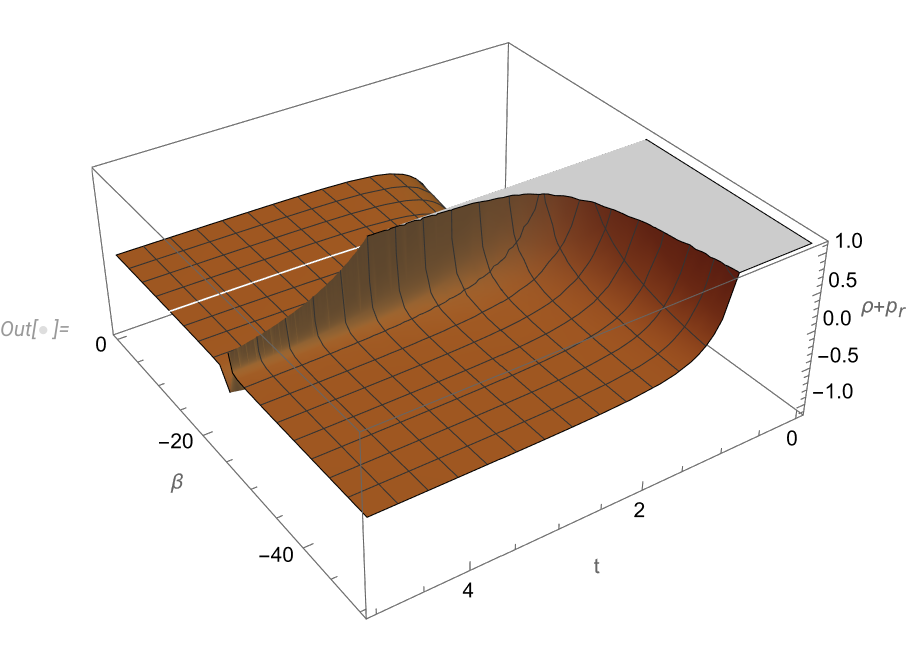}}}
	\caption{Plots demonstrating the variation of $\rho+p_r$ with parameter $\beta $ and time $t$ for $a)$ Model I and $b)$ Model II for power law scale factor. }
	\label{betavaluepl}
	\end{figure*}

\begin{figure*}[!]
		\centering
		\subfloat[]{{\includegraphics[width=8cm]{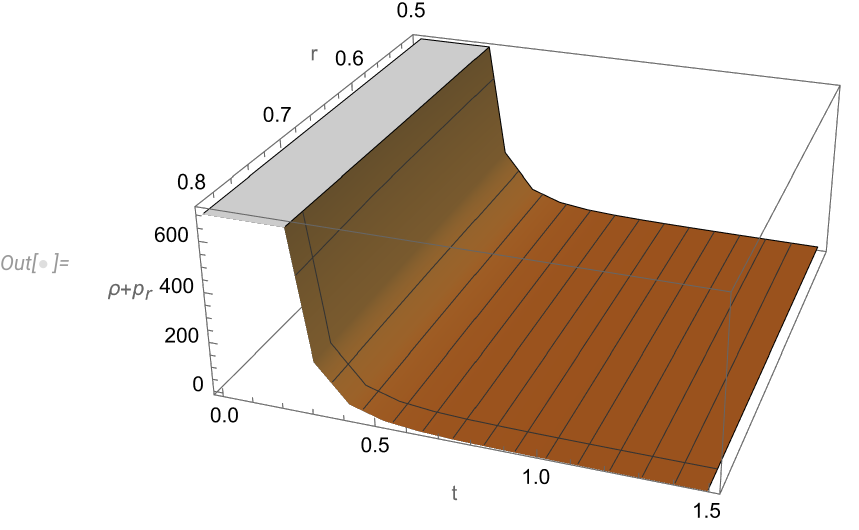}}}\qquad
		\subfloat[]{{\includegraphics[width=8cm]{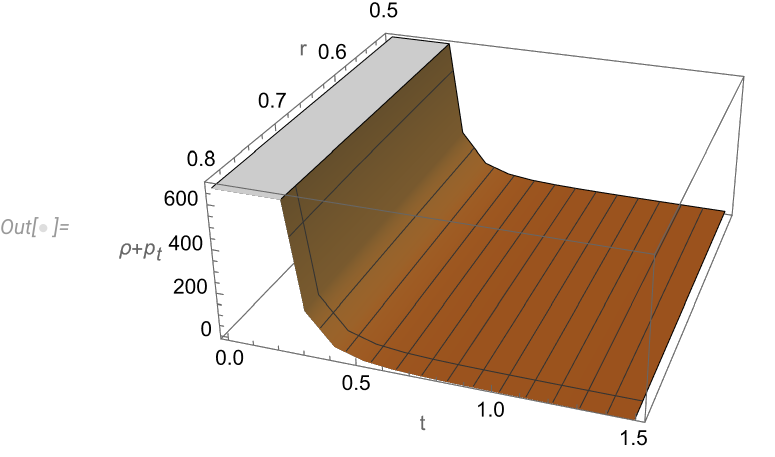}}}
		\subfloat[]{{\includegraphics[width=8cm]{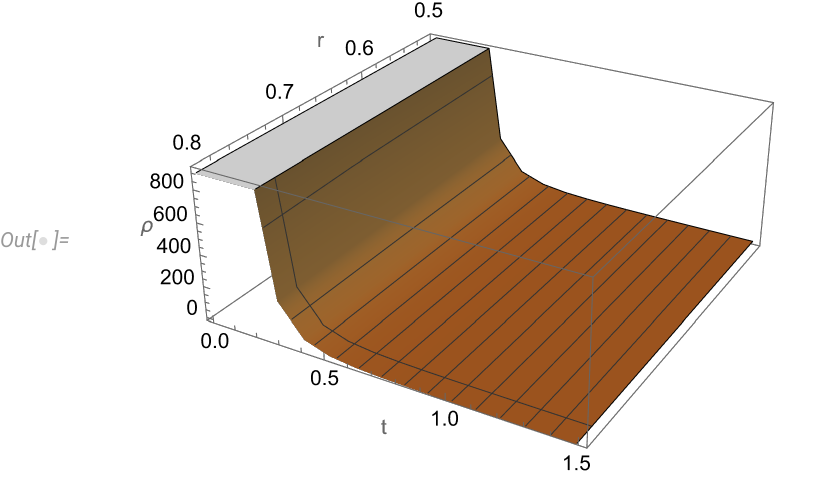}}}\\
		\subfloat[]{{\includegraphics[width=8cm]{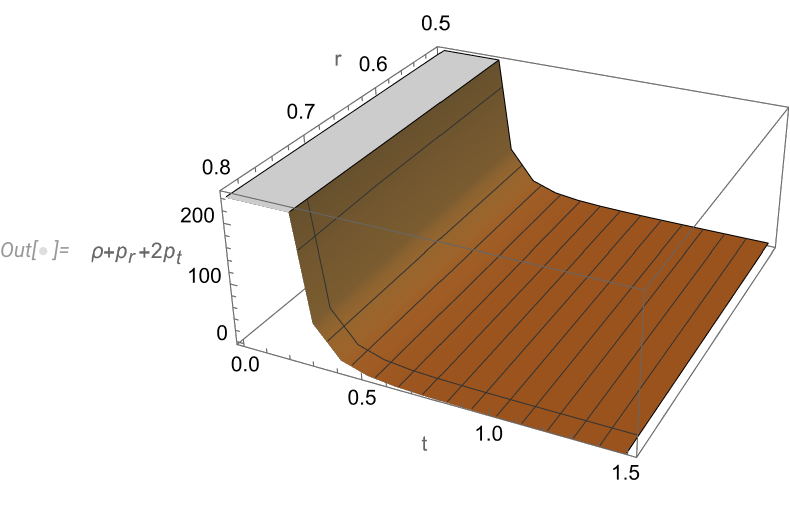}}}	
		\subfloat[]{{\includegraphics[width=8cm]{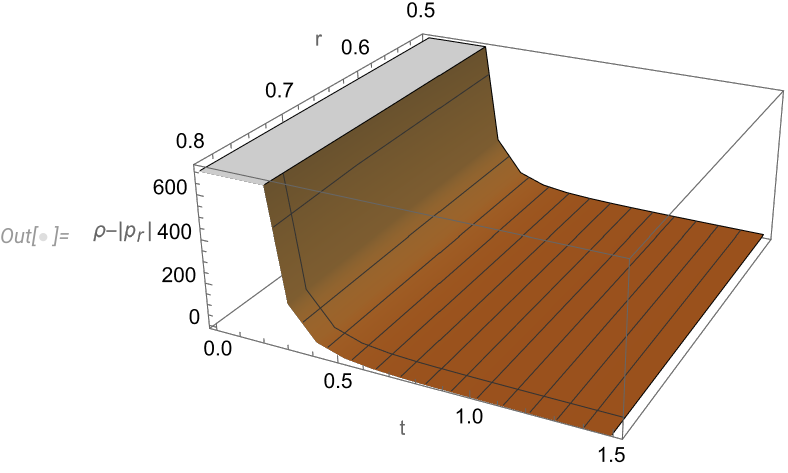}}}\qquad
		\subfloat[]{{\includegraphics[width=9cm]{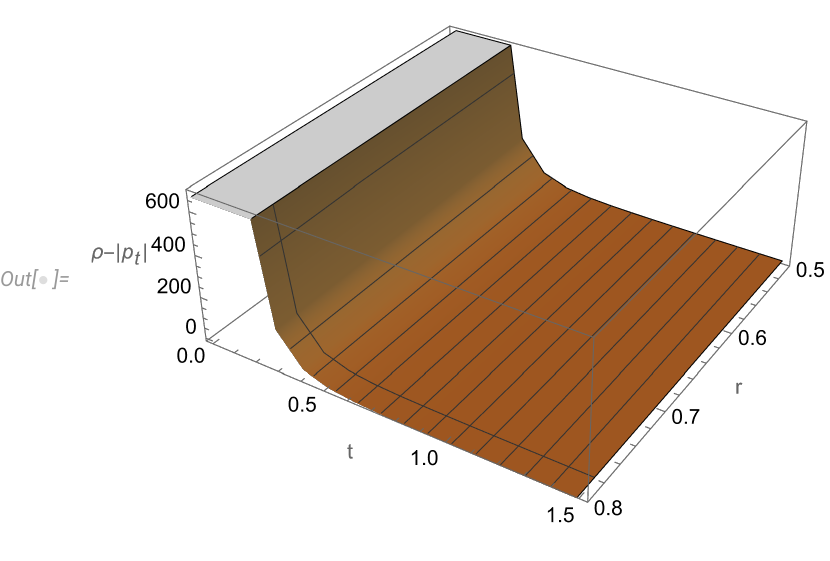}}}\\
		\caption{Plots demonstrating the variation of energy condition components  with radial distance $ r $ and time $t$ for Model I considering power law scale factor}. The nature of the plots are obtained considering $ \alpha=1, ~ \beta=-32,~a_0=2.5,~n=0.66 ,m=2$ and $ r_0=0.5 $(throat~ radius).
		\label{en_con_plot1}
	\end{figure*}

\begin{figure*}[!]
		\centering
		\subfloat[]{{\includegraphics[width=8cm]{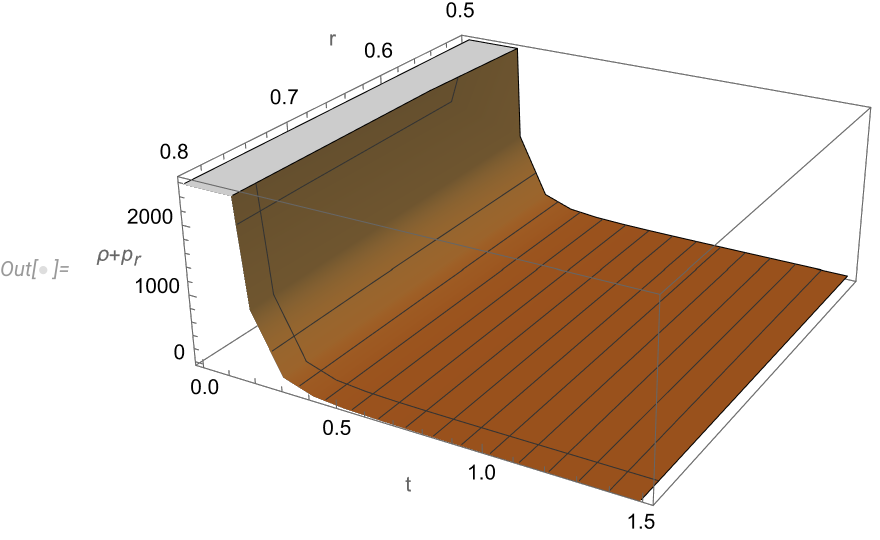}}}\qquad
		\subfloat[]{{\includegraphics[width=8cm]{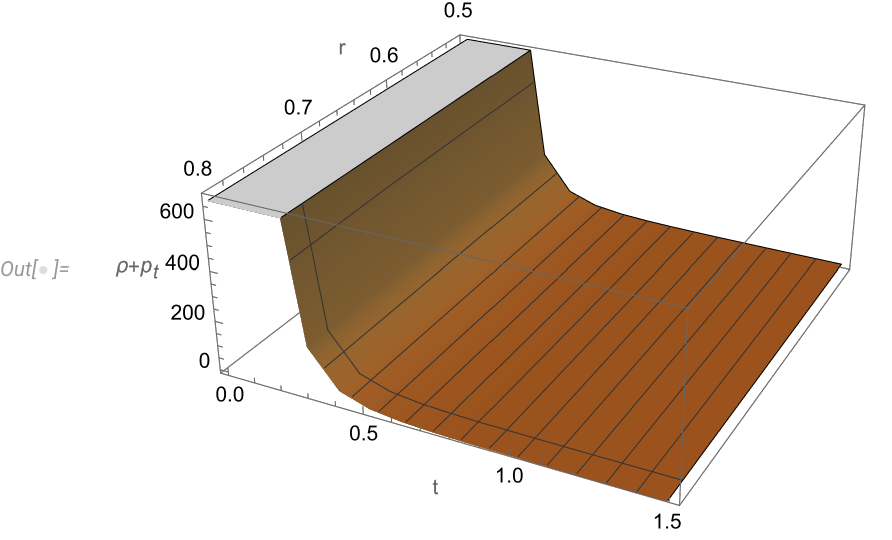}}}
		\subfloat[]{{\includegraphics[width=8cm]{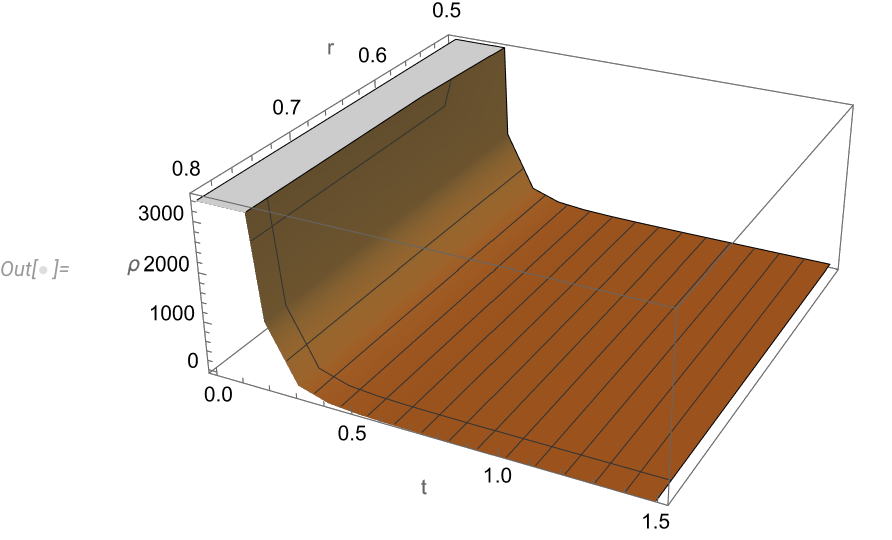}}}\\
		\subfloat[]{{\includegraphics[width=8cm]{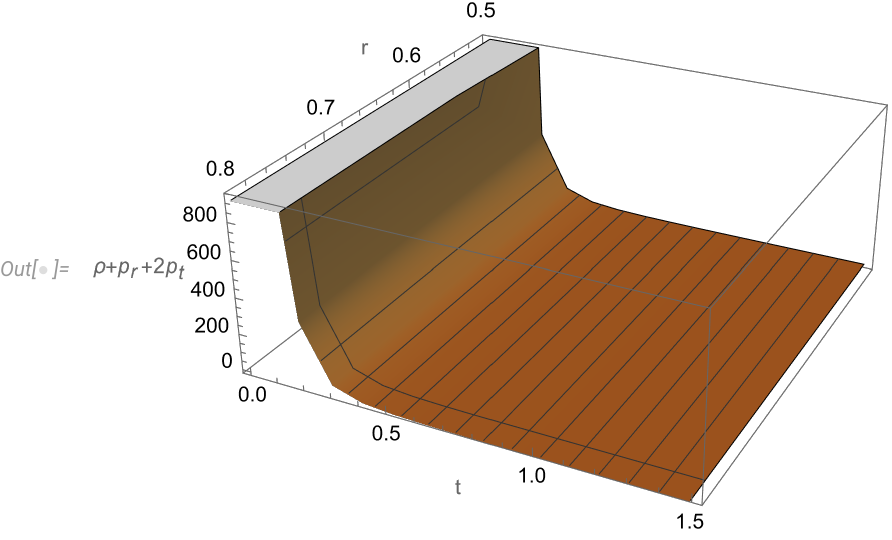}}}	\subfloat[]{{\includegraphics[width=8cm]{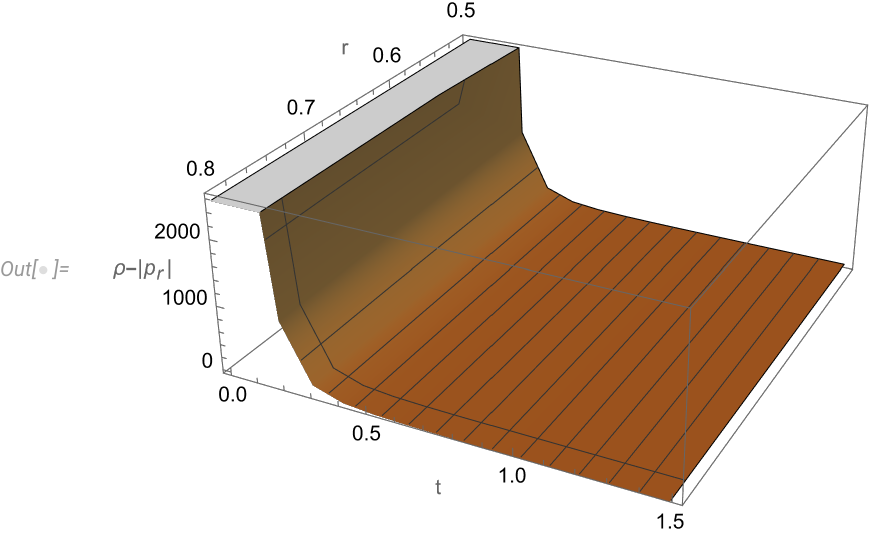}}}\qquad
		\subfloat[]{{\includegraphics[width=9cm]{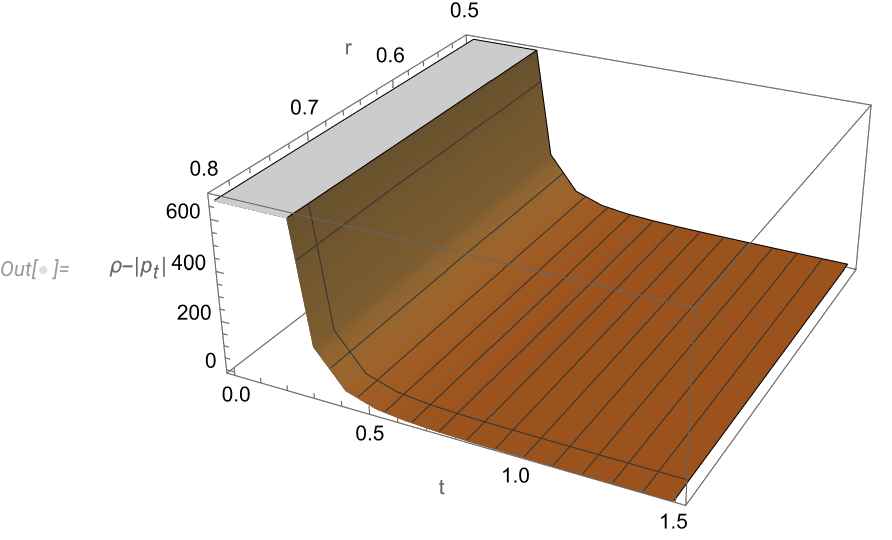}}}\\
		\caption{Plots demonstrating the variation of energy condition components  with radial distance $ r $ and time $t$ for Model II considering power law scale factor. The nature of the plots are obtained considering $ \alpha=1, ~ \beta=-32,~a_0=2.5,~n=0.66,m=2 $ and $ r_0=0.5 $(throat~ radius).}
		\label{en_con_plot2}
	\end{figure*}

For power law form of the scale factor the choices of the parameters are $ \alpha=1, ~ \beta=-32,~a_0=2.5,~n=0.66, m=2  $ and $ r_0=0.5$(throat~ radius). The negative value of $\beta$ can be attributed to the plots given by figure (\ref{betavaluepl})  for which the values of $\rho+p_r$ is positive at the throat. For Model I and Model II the plots of the energy condition components are given by the figures(\ref{en_con_plot1}) and (\ref{en_con_plot2}) respectively. It is observed that all the energy conditions are satisfied for both the models for the chosen parameters, providing the possibility for traversable evolving wormholes to exist with ordinary matter at the throat.
\\

$\bullet$ \textbf{Exponential Scale Factor: }

\begin{figure*}[h!]
	\centering
	\subfloat[]{{\includegraphics[width=8cm]{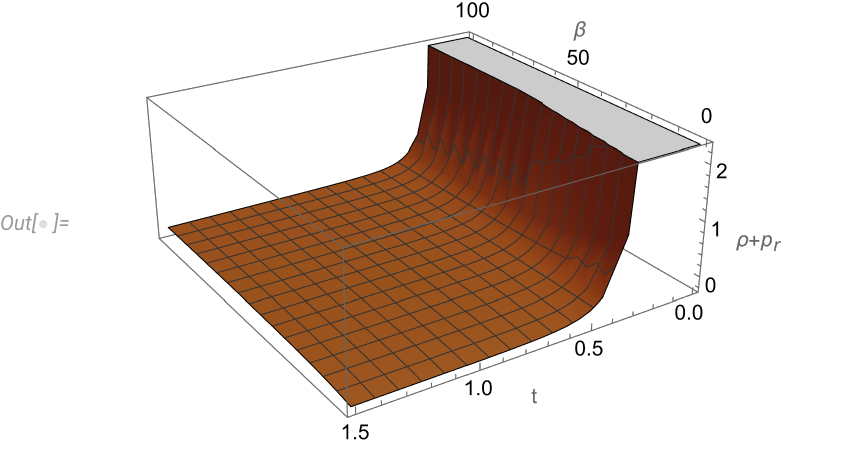}}}\qquad
	\subfloat[]{{\includegraphics[width=8cm]{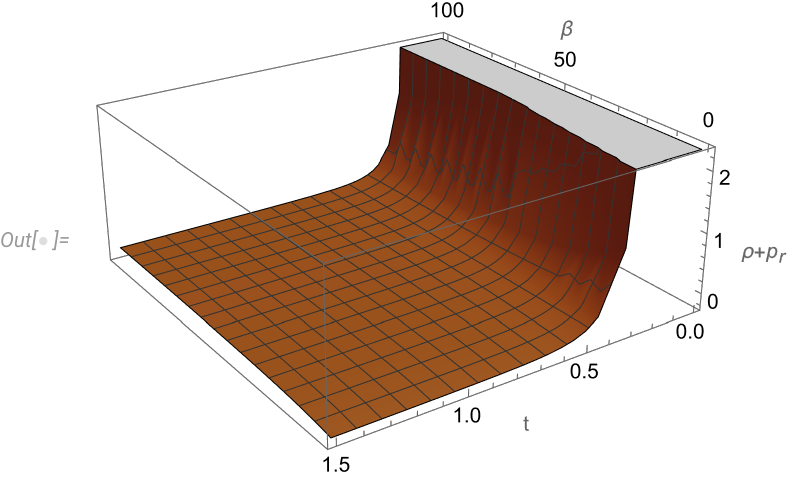}}}
	\caption{Plots demonstrating the variation of $\rho+p_r$ with parameter $\beta $ and time $t$ for $a)$ Model I and $b)$ Model II for exponential scale factor. }
	\label{betavalueexp}
\end{figure*}

\begin{figure*}[h!]
	\centering
	\subfloat[]{{\includegraphics[width=8cm]{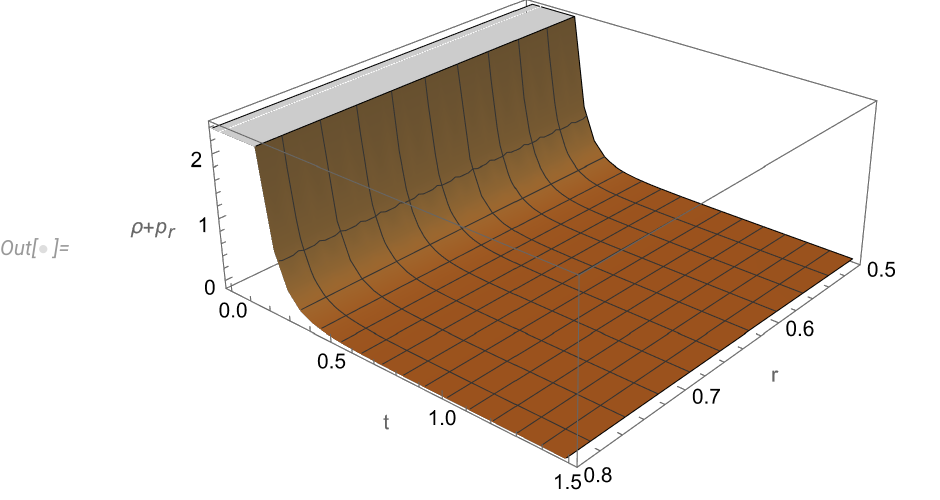}}}\qquad
	\subfloat[]{{\includegraphics[width=8cm]{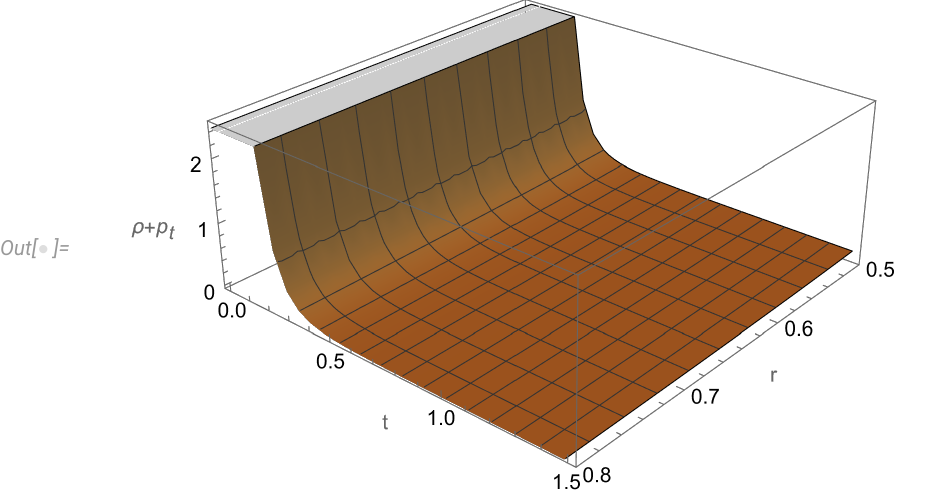}}}
	\subfloat[]{{\includegraphics[width=8cm]{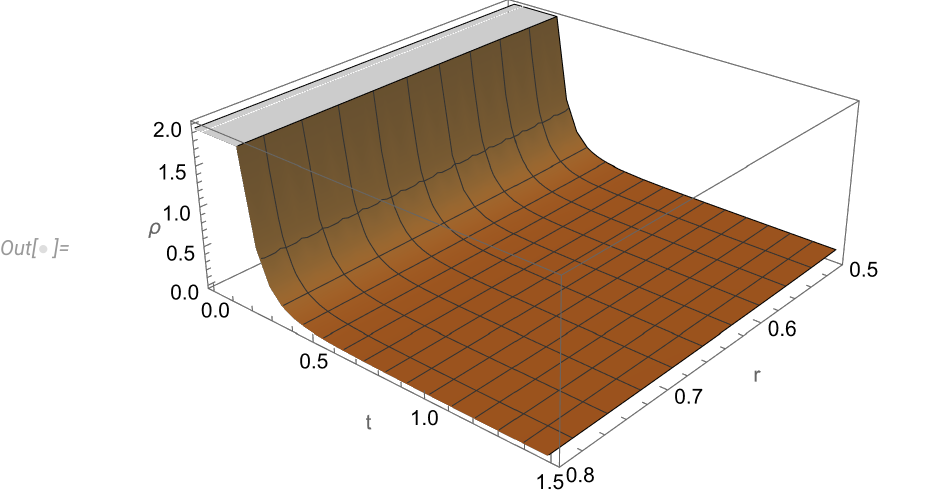}}}\\
	\subfloat[]{{\includegraphics[width=8cm]{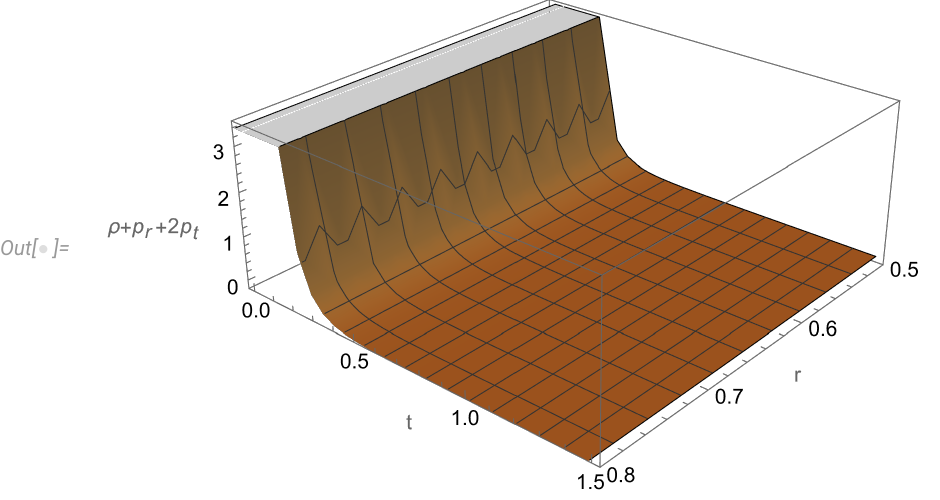}}}	
	\subfloat[]{{\includegraphics[width=8cm]{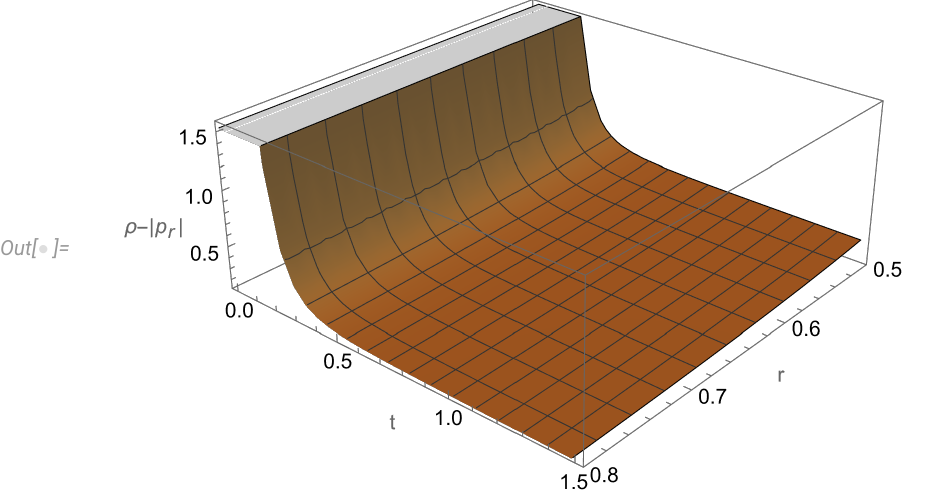}}}\qquad
	\subfloat[]{{\includegraphics[width=9cm]{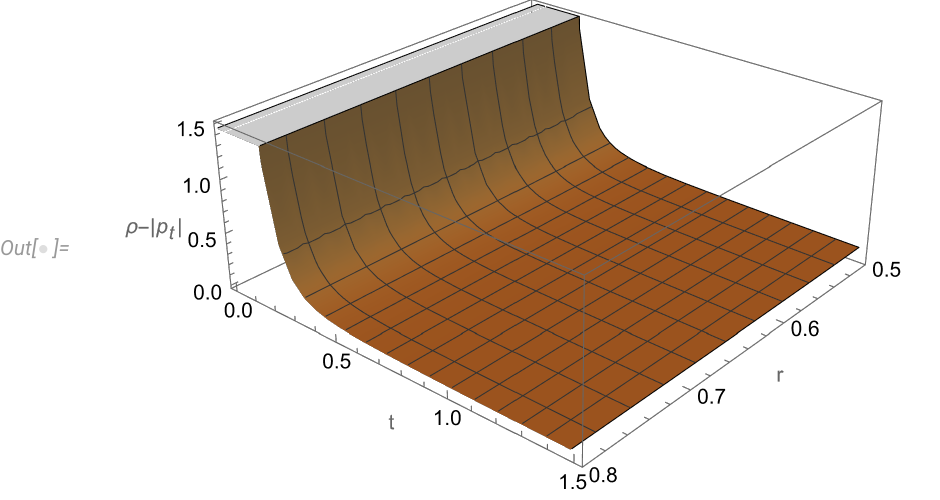}}}\\
	\caption{Plots demonstrating the variation of energy condition components  with radial distance $ r $ and time $t$ for Model I considering exponential form of scale factor. The nature of the plots are obtained considering  $ \alpha=1, ~ \beta=50,~a_0=2.3,~n=0.3 $, $\mu=0.2$, $m=2$ and $ r_0=0.5$(throat~ radius).}
	\label{en_con_plot4}
\end{figure*}

\begin{figure*}[!]
	\centering
	\subfloat[]{{\includegraphics[width=8cm]{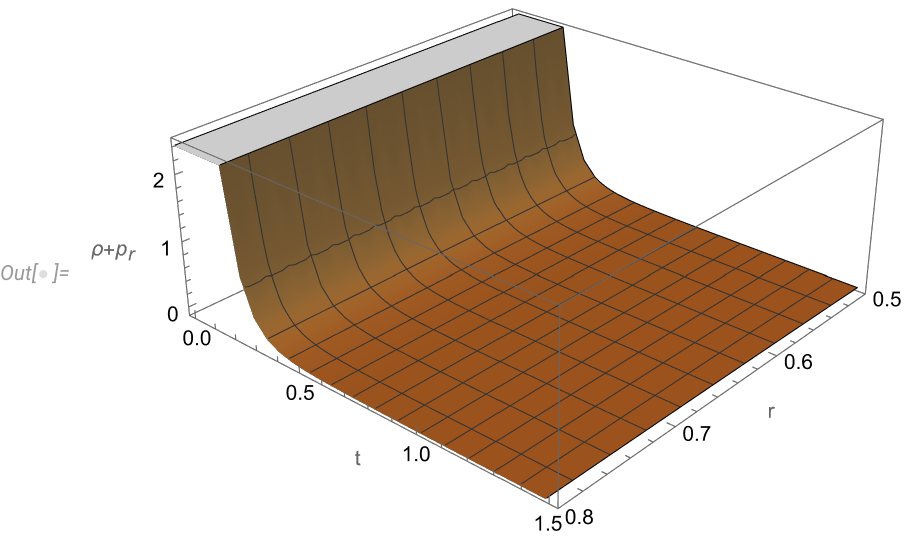}}}\qquad
	\subfloat[]{{\includegraphics[width=8cm]{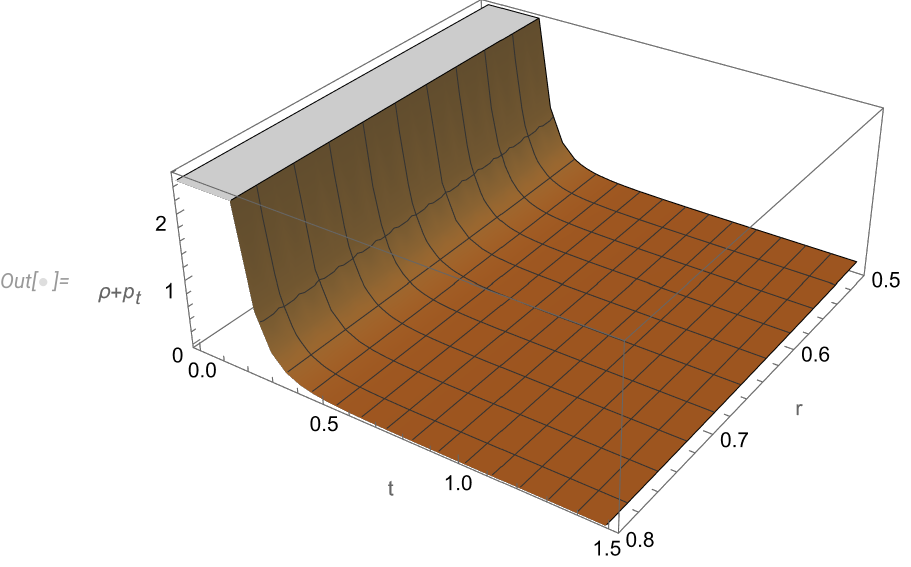}}}
	\subfloat[]{{\includegraphics[width=8cm]{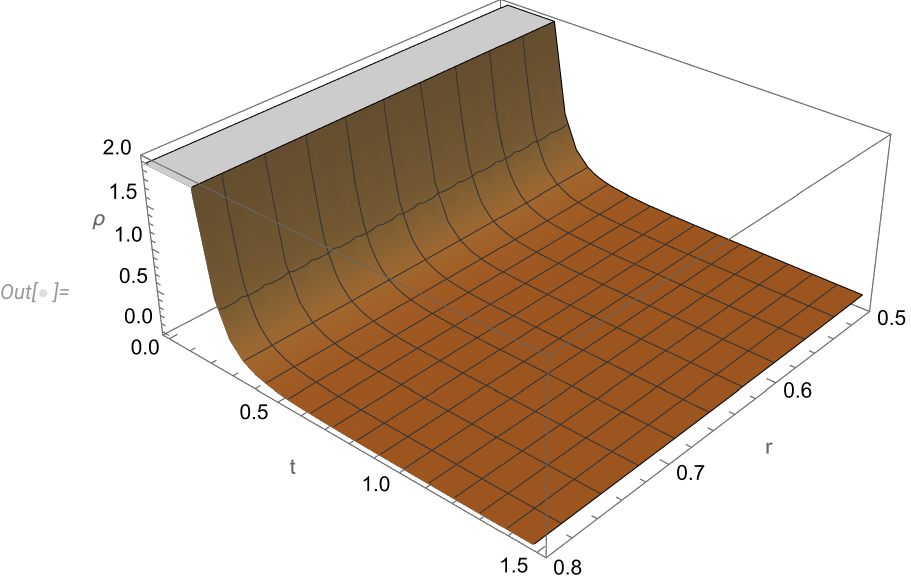}}}\\
	\subfloat[]{{\includegraphics[width=8cm]{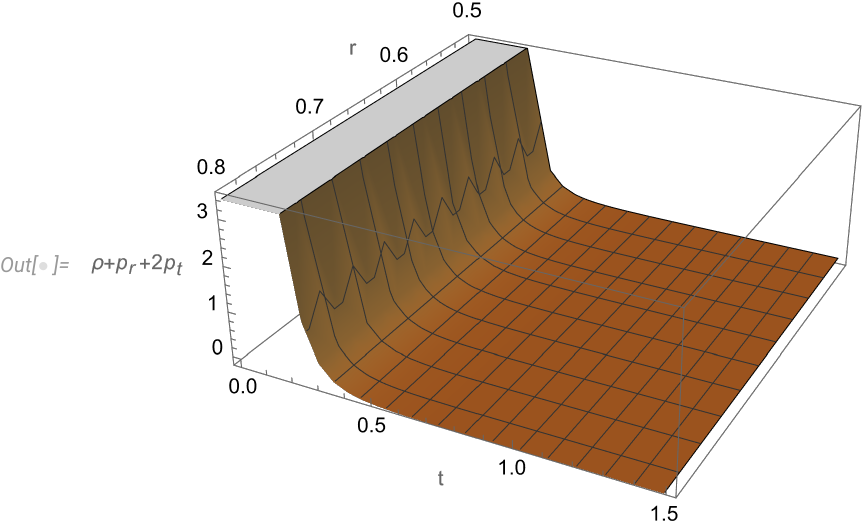}}}	\subfloat[]{{\includegraphics[width=8cm]{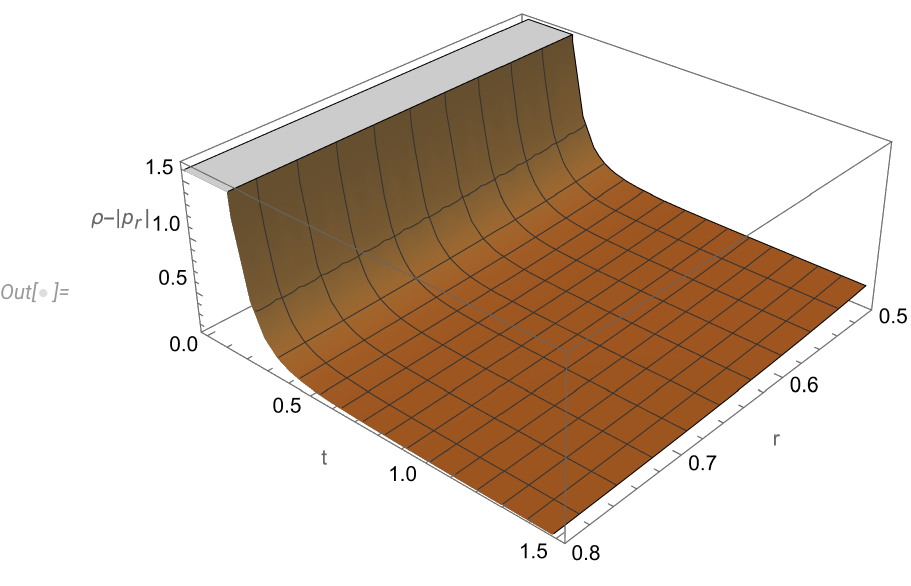}}}\qquad
	\subfloat[]{{\includegraphics[width=8cm]{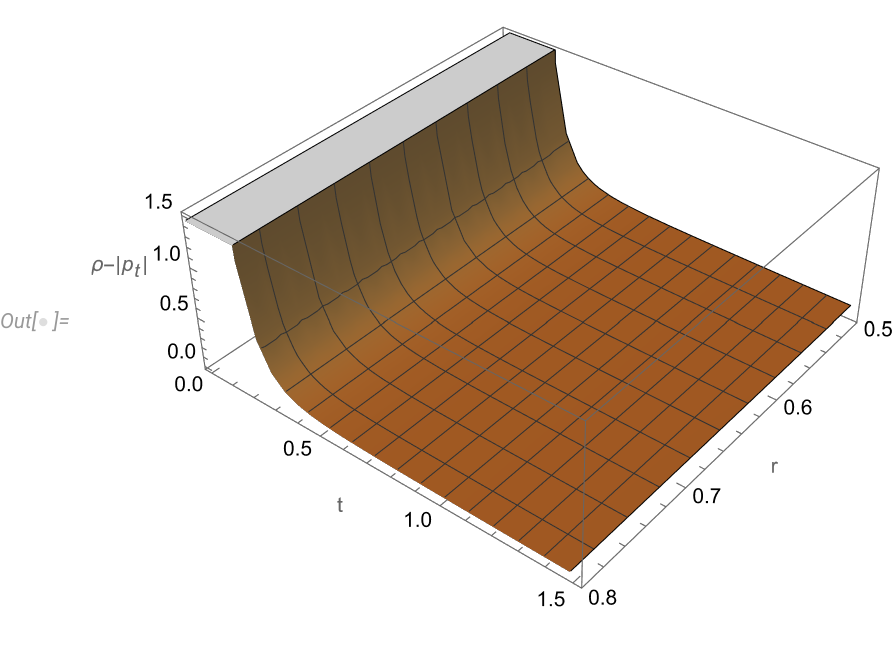}}}\\
	\caption{Plots demonstrating the variation of energy condition components  with radial distance $ r $ and time $t$ for Model II considering exponential form of scale factor. The nature of the plots are obtained considering  $ \alpha=1, ~ \beta=50,~a_0=2.3,~n=0.3 $, $\mu=0.2$, $m=2$ and $ r_0=0.5$(throat~ radius).}
	\label{en_con_plot5}
\end{figure*}

For the exponential choice of the scale factor, the choices of the parameters are $ \alpha=1, ~ \beta=50,~a_0=2.5,~n=0.3 ,\mu=0.2 ,m=2 $ and $ r_0=0.5$(throat~ radius). The  value of $\beta$ can be attributed to the plots given by figure (\ref{betavalueexp})  for which the values of $\rho+p_r$ is positive at the throat. For Model I and Model II the plots of the energy condition components are given by the figures(\ref{en_con_plot4}) and (\ref{en_con_plot5}) respectively.Here, also, it is observed that all the energy conditions are satisfied for both the models for the chosen parameters, providing the possibility for traversable evolving wormholes to exist with ordinary matter at the throat.


	

	The singularity and the horizon problem of the big bang cosmology has gathered significant attention over the past years. In order to provide an alternative model bypassing this issue Ellis and Marteens put forward a cosmological model called an Emergent Universe Model\cite{Ellis:2002we}. An emergent universe represents a model in which a time-like singularity is absent, showing an almost static state in the remote past (as \( t \to -\infty \)). Gradually, the model evolves into an inflationary phase. This view reflects a modern reinterpretation and extension of the classical Lemaître-Eddington universe framework.This nonsingular cosmological model is rooted in Einstein's static universe and provides solutions to horizon-related issues. The emergent scenario exhibits features common to both inflationary and bouncing cosmologies. Similar to inflationary cosmology, it begins with a hot, compact universe. However, like bouncing cosmology, it spans time from negative to positive infinity, maintaining a non-singular evolutionary process. 
	The universe is said to be emergent if it satisfies certain conditions so as to avoid the big bang singularity. The conditions for emergent scenario are\cite{Chakraborty:2014ora}-\cite{Dutta:2016kkl}:
	\begin{itemize}\label{EUCON}
		\item $a\rightarrow a_0$, $H \rightarrow 0$ as $t \rightarrow -\infty$.
		\item  $a \simeq a_0$, $H \simeq 0$, $t << t_0$.
	\end{itemize}
Hence, from (\ref{sf eu}), we see that as 
$t \rightarrow -\infty(\mu>0)$, then $a\rightarrow a_0$. Also the Hubble parameter becomes vanishingly small as $t \rightarrow -\infty(\mu>0)$ and satisfies the conditions necessary for the universe to be emergent. This means that in the infinite past the universe was eternally as Einstein's static universe. Subsequently, the universe starts expanding and gradually enters in inflationary phase, thereby the big bang singularity may be avoided.

In the present context, this means that the present evolving wormhole was eternally in the Einstein's static era, and subsequently the wormhole starts evolving into the inflationary era of expansion.

		\begin{figure*}[h!]
		\centering
		\subfloat[]{{\includegraphics[width=8cm]{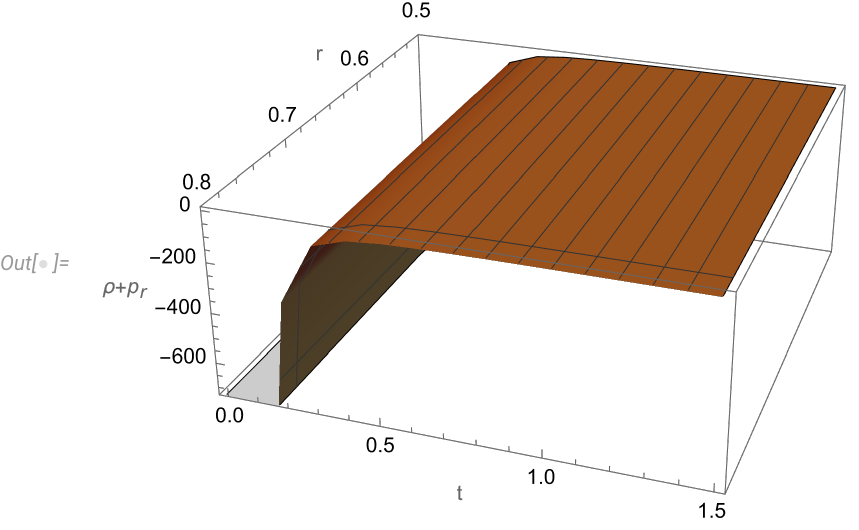}}}
		\subfloat[]{{\includegraphics[width=8cm]{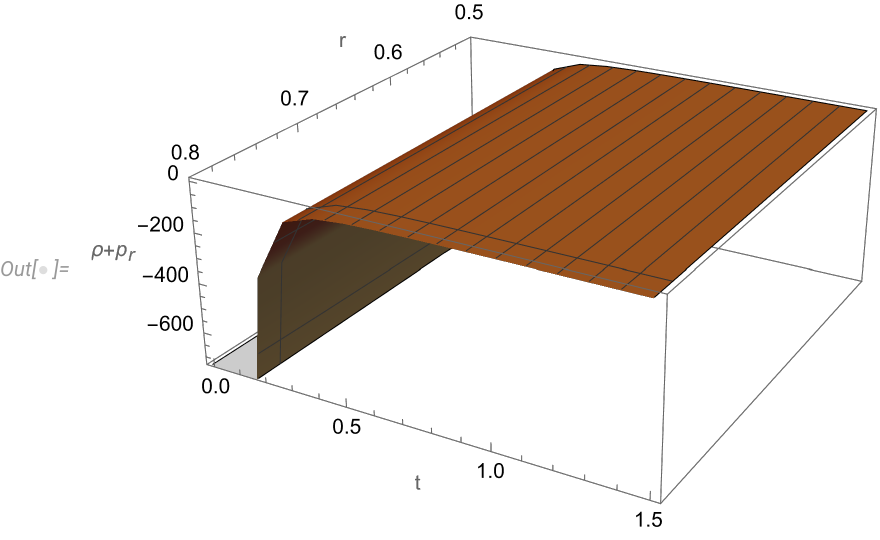}}}\\
		\subfloat[]{{\includegraphics[width=8cm]{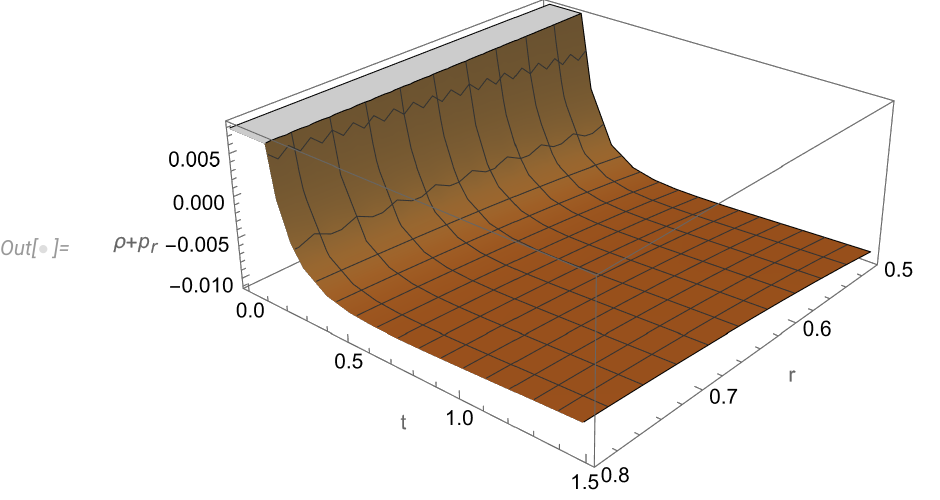}}}
		\subfloat[]{{\includegraphics[width=8cm]{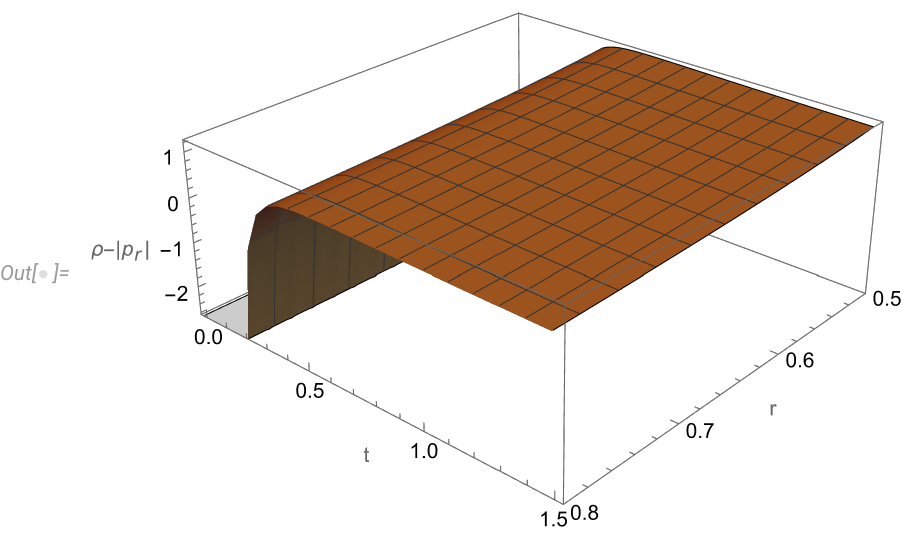}}}	
	
		\caption{Plots demonstrating the violation of NEC with radial distance $ r $ and time $t$ for Model I and Model II for power law scale factor (a and b) and for exponential scale factor (c and d) considering $\beta=0$ violation of NEC and DEC for Model I and Model II respectively are shown. }
		\label{betavaluezero}
	\end{figure*}
	
	\section{Results and Discussions}\label{discuss}

	In the present work we have obtained an evolving wormhole model in the realm of $f(R,T)$ gravity. Due to the very complicated and coupled form of the field equations we  assume (i) the redshift function $\phi=0$ and (ii) $f(R,T)=\alpha R^m+\beta T$. For wormhole solution we have chosen two shape functions(which are well known in the literature) and are named as Model I and  Model II. respectively. For the scale factor $a(t)$ we have restricted ourselves to (i) power law form  (ii) exponential form. Subsequently, a wormhole solution is obtained and the energy conditions namely the NEC, WEC, SEC and DEC are investigated. Regarding the matter content in the current wormhole solutions, it is noteworthy that, unlike much of the existing \( f(R, T) \) wormhole literatures(discussed above), no specific equation of state (EoS) has been assumed in this study. Given the limited understanding of the wormhole EoS, this feature represents a significant advantage of the solutions presented here. It is important to note that with potential future detections of wormholes, it may become possible to place constraints on their equations of state. 
	
	In this gravity model the parameter $\alpha$ has to be greater than zero to imply cosmic inflation. The inclusion of  matter-geometry coupling term $\beta T$  is done with the purpose of considering quantum effects in a gravity theory, or even the existence of imperfect fluids in the universe which can resolve the issues of wormhole matter violating the NEC, WEC and SEC in GR\cite{Harko:2011kv}. The non zero value of the parameter $\beta$ and  consideration of the $f(R,T)$ gravity theory can also be judged by the fact that for the zero values of $\beta$ there is violation of the energy conditions for all the cases. The plot given by figure (\ref{betavaluezero}) validates the motivation behind considering this gravity theory.
	
The investigation brings forth interesting results suggesting evolving traversable wormhole geometry can be achieved within \( f(R, T) \) theory without requiring exotic matter that violates the NEC. In both the cases of the scale factor the NEC is satisfied by both the shape function.However, in \( f(R, T) \) gravity, the NEC alone is no longer considered a significant accomplishment in this theory. Results considering static wormhole geometry satisfying all energy conditions can be found in the literature\cite{Rosa:2022osy}. For the two shape functions considering the power law form of the scale factor form plots (\ref{en_con_plot1}) and (\ref{en_con_plot2}), it can be concluded that not only NEC but all the energy conditions are satisfied for the given values. From the plots (\ref{en_con_plot4}) and (\ref{en_con_plot5}) we see that the shape functions in the exponential form of scale factor satisfies all the energy conditions.  The energy conditions are satisfied at the throat of the wormhole, which allows for non-exotic matter to thread the wormhole throat geometry, results from additional term `\( \beta T \)'. The additional degrees of freedom may allow for wormholes to be supported by non-exotic matter EoS. The $T$ dependence in \( f(R, T) \) theory might mark initial steps toward incorporating quantum effects within a gravitational framework\cite{Harko:2011kv}. Such a perspective, absent in General Relativity, could clarify how energy conditions are fulfilled in these models.

 As previously mentioned, wormholes supported by non-exotic matter are complex constructs, and the material correction term predicted by \( f(R, T) \) theory, potentially linked to the presence of imperfect fluids in the cosmos, may significantly advance our understanding of wormholes.Through alternative gravity theories or by considering unconventional EoS several cosmology and astrophysics the challenges are being addressed. For example, the problem of dark energy, massive pulsars, the present universe experiencing an accelerated expansion phase and many others\cite{WMAP:2012nax, Starobinsky:2007hu, Demorest:2010bx, Moraes:2015uxq,Fortin:2014mya}.
These concepts may be applied to wormholes, as they too are exotic astrophysical objects.
Hence, having a solution of an evolving wormhole in this  modified theory of gravity which explains the acceleration of the universe also accelerates the understanding of how exotic objects like wormholes evolve with the expansion of the universe. 
Finally, we may also conclude, the exponential choice of the scale factor for the present dynamical wormhole scenario has a singularity free emergent configuration at infinite past.
\par 
The present work shows that for $f(R,T)$ modified gravity theory it is possible to have evolving wormhole without considering any exotic matter. Thus for $f(R,T)$ gravity theory similar to static wormhole geometry the evolving wormhole exists with normal matter which indicates that the exotic nature of the matter component may arrive due to the geometric matter part of the present modified gravity theory.
	

\end{document}